\documentclass[a4paper,fleqn,usenatbib]{mnras}
\usepackage[T1]{fontenc}
\usepackage{ae,aecompl}
\usepackage{graphicx}
\usepackage{amssymb}
\usepackage{amsmath}
\usepackage{mathptmx}
\usepackage{epstopdf}
\usepackage{booktabs}
\usepackage{float}
\usepackage{subfig}
\usepackage{epsfig,color,ulem}
\usepackage{tabularx}

\newcommand{\A}{{\it A}}
\newcommand{\B}{{\it B}}
\newcommand{\C}{{\it C}}
\newcommand{\D}{{\it D}}
\newcommand{\E}{{\it E}}
\newcommand{\F}{{\it F}}
\newcommand{\msun}{\,{\rm M_{\odot}}}
\newcommand{\cm}{\,{\rm cm}}

\newcommand{\erg}{\,{\rm erg}}
\newcommand{\tobs}{\,{t_{\rm obs}}}
\newcommand{\ts}{\,{t_{\rm sim}}}

\newcommand{\rph}{\,{r_{\rm ph}}}

\newcommand{\s}{\,{\rm s}}
\def\gsim{ \lower .75ex \hbox{$\sim$} \llap{\raise .27ex \hbox{$>$}} }
\def\lsim{ \lower .75ex\hbox{$\sim$} \llap{\raise .27ex \hbox{$<$}} }
\def\app#1#2{%
	\mathrel{%
		\setbox0=\hbox{$#1\sim$}%
		\setbox2=\hbox{%
			\rlap{\hbox{$#1\propto$}}%
			\lower1.1\ht0\box0%
		}%
		\raise0.25\ht2\box2%
	}%
}

\title[High efficiency photospheric emission in GRBs]{High efficiency photospheric emission entailed by formation of a collimation shock in gamma-ray bursts}
\author[Gottlieb, Levinson \& Nakar]{
	Ore Gottlieb\thanks{oregottlieb@mail.tau.ac.il},
	Amir Levinson,
	Ehud Nakar
	\\
	{The Raymond and Beverly Sackler School of Physics and
		Astronomy, Tel Aviv University, Tel Aviv 69978, Israel}
}
\pubyear{2019}
\begin{document}
	\label{firstpage}
	\pagerange{\pageref{firstpage}--\pageref{lastpage}}
	\maketitle	
	\begin{abstract}
		The primary dissipation mechanism in jets of gamma-ray bursts (GRBs), and the high efficiency of the prompt emission are long standing issues. One possibility is strong collimation of a weakly magnetized relativistic jet by the surrounding medium, which can considerably enhance the efficiency of the photospheric emission. We derive a simple analytic criterion for the radiative efficiency of a collimated jet showing that it depends most strongly on the baryon loading. We confirm this analytic result by 3D numerical simulations, and further find that mixing of jet and cocoon material at the collimation throat  leads to a substantial stratification of the outflow as well as sporadic loading, even if the injected jet is uniform and continuous. One consequence of this mixing is a strong angular dependence of the radiative efficiency. Another is large differences in the Lorentz factor of different fluid elements that lead to formation of internal shocks. Our analysis indicates that in both long and short GRBs a prominent photospheric component cannot be avoided when observed within an angle of a few degrees to the axis, unless the asymptotic Lorentz factor is limited by baryon loading at the jet base to $\Gamma_\infty <100$ (with a weak dependence on outflow power). Photon generation by newly created pairs behind the collimation shock regulates the observed temperature at $\sim 50~\theta_0^{-1}$ keV, where $\theta_0$ is the initial jet opening angle, in remarkable agreement with the observed peak energies of prompt emission spectra. Further consequences for the properties of the prompt emission are discussed at the end.	
\end{abstract}
	
	\section{Introduction}
	\label{sec:introduction}	
	The origin of the prompt emission in gamma-ray bursts (GRBs) is a long standing issue.   The conventional wisdom has been 
	that the observed gamma ray flash is produced in a relativistic jet, following its emergence from the  dense medium  
	surrounding the central engine - the envelope of the progenitor star in long GRBs, and the neutron star merger  ejecta in short GRBs.   Within the framework of this model two central questions arise:
	(i) what determines the radiative efficiency during the prompt emission phase, which in long GRBs 
	at least appears to be rather high (e.g., \citealt{Lloyd2004,racusin2011}, but see \citealt{santana2014,xiang2015} for a recent analysis); and (ii) what is the origin of the non-thermal spectrum observed in most bursts.    
	
	High efficiency can be naturally achieved if the photosphere of the emitting outflow is located beneath the coasting radius.
	In such a case, one might naively expect a nearly black body spectrum,
	as predicted in the original fireball model \citep{Paczynski1986,Goodman1986}.
	However, as demonstrated in several works (e.g. \citealt{Giannios2008,Beloborodov2010,Lazzati2010,Meszaros2011,Veres2012,Vurm2013,keren2014}), dissipation below the photosphere, at a modest optical depth, can give 
	rise to a significant broadening of the emitted spectrum that would make it appear non-thermal. 
	Hence, photospheric emission does not necessarily entail a thermal spectrum!  Whether sub-photospheric dissipation 
	can account for the spectrum observed in the majority of sources is yet an open question.  
	If it would turn out that it cannot, and that the emission needs to be produced well above
	the photosphere by some mechanism (e,g., internal shocks or magnetic dissipation), it would mean that the photosphere must be located well above the coasting radius in order to avoid an excessive photospheric component.
	Thus, the location of the photosphere is a key issue in GRB theory.

	Quite generally, the location of the photosphere  depends on the outflow injection radius, its baryon load  and its isotropic equivalent power.
	In a purely conical hydrodynamic outflow injected from the vicinity of the compact object ($R_0\sim10^7$ cm),
	and having a power compatible with a typical GRB luminosity,
	the photosphere will be located within the acceleration zone if the baryon
	rest energy flux, $\dot{M}c^2$, comprises less than a promille of the total outflow power
	(implying an upper limit on the asymptotic Lorentz factor in excess of $10^3$).  
	If the baryon load exceeds this value, photospheric emission should be strongly 
	suppressed, unless a significant fraction of the outflow bulk energy dissipates just below the photosphere.  
	However, in practice the outflow is not purely conical but rather strongly collimated  by the surrounding medium,
	and this, as will be shown below, can give rise to a considerable enhancement of the efficiency of photospheric emission. 
	Specifically, we find that a significant photospheric component cannot be avoided unless the asymptomatic Lorentz factor  is limited to $\Gamma_\infty\lsim 100$.

	The effect of collimation on the efficiency of photospheric emission was identified for the first time in 2D numerical simulations by \cite{Lazzati2009}, and later also in 3D simulations \citep{Lopez-Camara2013,Ito2015}. Those earlier studies were restricted to specific conditions (very powerful outflows, long engine time and a specific progenitor type) and/or numerical limitations, and the question remains how this efficiency scales with the burst properties.  In this paper we derive a simple analytic criterion for the critical load of a 
	strongly collimated hydrodynamical outflow below which efficient 
	photospheric emission is expected.   We compare this formula with the results of 3D relativistic-hydrodynamic (RHD) simulations
	and find a remarkably good agreement. We also find sporadic loading of streamlines by mixing of jet and cocoon material
	in the vicinity of the collimation shock. Our simulations indicate stratification of the flow by this additional loading, that renders the radiative efficiency angle dependent.  For typical GRB parameters, the efficiency is
	found to be exceptionally high within a core of opening angle of about one third of the opening angle at injection, and it declines with increasing inclination outside the core. We conclude that if GRBs are indeed highly relativistic as commonly perceived, photospheric emission should dominate the prompt emission if the outflow becomes weakly magnetized before reaching the photosphere. We also find that the rest-frame temperature just behind the collimation shock is robustly maintained at about $50$ keV by a pair production thermostat, and show that this provides a natural explanation for the observed peak energy of the prompt emission spectrum. Further implications 
	for the properties of the observed spectrum are discussed at the end.
	
	Photospheric emission has been discussed in many earlier works (e.g., \citealt{eichler2000,ryde2004,peer2005,deng2014,beloborodov2017}).  We emphasize that the focus of this paper is not on explaining the properties of the prompt emission by a photospheric model, but rather to show that formation
	of a collimation shock robustly necessitates high efficiency photospheric emission from first principles.  Any model that invokes a hydrodynamic 
	(or a weakly magnetized) outflow, including conversion of Poynting flux jet well inside the star, must account for this fact.  Conversion of a Poynting
	jet outside the star but below the photosphere constitutes a seperate case that needs a different treatment.

	\section{Analytic Criterion for minimum efficiency}
	\label{sec:analytic}
\begin{figure*}
		\centering
		\includegraphics[scale=0.58,trim=0.4cm 2cm 0cm 0.5cm]{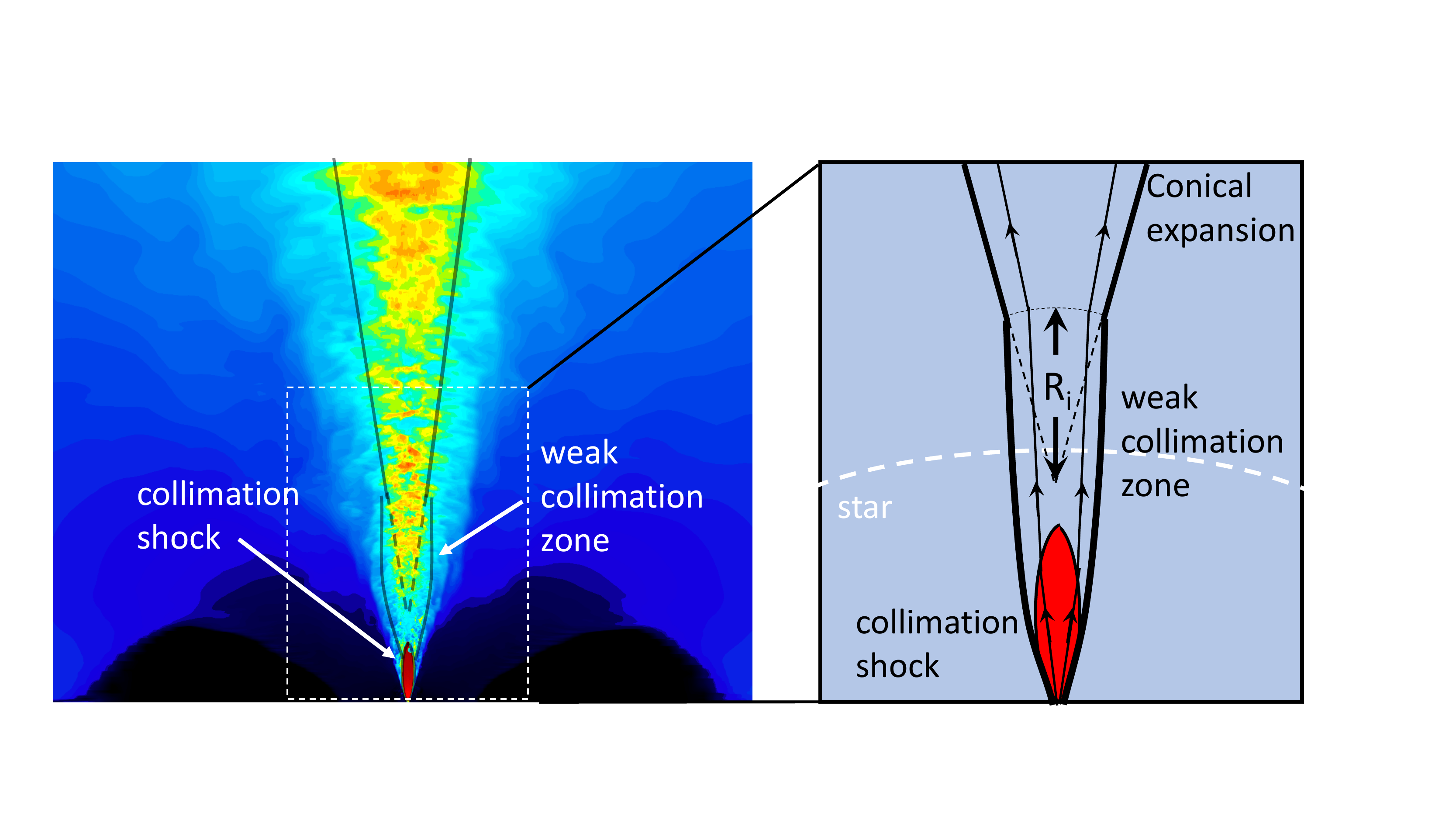}
		\caption[]{Left: A snapshot of the simulation at late time showing the primary collimation shock (red) and 
			the weak collimation zone, as indicated.  Right: 
			Schematic illustration of the inner region (the dashed white box in the left panel).  The weak collimation zone extends 
			up to about two stellar radii, above which the outflow re-expands conically. $ R_i $ is the distance
			between the base of the re-expanding conical flow and the focus of the radial streamlines.
		}
		\label{fig:sketch}
	\end{figure*}

	A strong collimation shock results from a supersonic deflection of streamlines by 
	the over-pressured cocoon that forms as the jet propagates through the dense envelope enshrouding the central engine.   
	The collimation shock propagates outwards as the outflow expands, reaching a relatively large radius by the time the outflow 
	breaks out of the dense medium.  It then continues to expand as the cocoon's pressure declines.  
	In typical long GRBs the collimation shock remains inside the progenitor's envelope,
	at a radius $r_s$ of the order of the stellar radius, for nearly the entire duration of the burst (see $ \S $ \ref{sec:simulation_results} below).  
	Over time, fresh fluid elements expelled from the central engine move along radial streamlines until reaching the collimation shock. 
	Upon crossing the  shock the streamlines are deflected and the fluid decelerates and heats up to relativistic temperatures. During the propagation of the flow inside the dense medium the external pressure is high and the jet is kept roughly cylindrical.  Once the flow emerges from the dense medium the confining pressure drops and the flow re-accelerates, quickly becoming conical again (see illustration in Figure \ref{fig:sketch}). 
	This re-acceleration of the flow from a large radius pushes the coasting radius closer to, and possibly above, the photosphere, thereby considerably increasing the radiative efficiency.  In short GRBs and exceptionally powerful long GRBs, the 
	collimation shock itself may break out following the jet and propagate at a mildly 
	relativistic speed to large distances, but the overall picture is qualitatively similar.
	
	To gain insight into the effect of the collimation shock, consider a conical, adiabatic outflow of opening angle $\theta_0$, 
	isotropic equivalent power $L_{jiso}=10^{52}L_{iso,52}$ ergs s$^{-1}$, and isotropic mass flux $\dot M_{iso}$.   
	Suppose that the outflow is injected from a radius  $R_0=10^7R_7$ cm, with an initial Lorentz factor $\Gamma=\Gamma_0\sim1$.
	From the flow equations it can be readily shown that $\eta_0=h\Gamma$ is conserved along streamlines,
	and that its value is given by $\eta_0=L_{jiso}/(\dot M_{iso} c^2)$.  It is essentially the maximum 
	Lorentz factor to which the flow 
	can accelerate before its energy is fully converted into bulk kinetic energy of the dragged baryons.
	Here  $h=1+4p/nm_pc^2$ is the dimensionless enthalpy per baryon,
	$p$ the pressure, and $n$ the proper baryon density.
	The photosphere will be located exactly at the coasting radius,  $r_{coast}\simeq\eta_0 R_0/\Gamma_0$,
	if $\eta_0=\eta_c$, where
	\begin{equation}
	\eta_c=\left(\frac{\sigma_T L_{jiso}\Gamma_0}{4\pi
		R_0m_pc^3}\right)^{1/4}=10^3~L_{iso,52}^{1/4}R_7^{-1/4}\Gamma_0^{1/4}, \label{eq:eta_c-def1}
	\end{equation}
	and $\sigma_T$ is the Thomson cross section \citep{Grimsrud1998}.
	If $\eta_0<\eta_c$ the photospheric radius is located above the coasting radius, and is given by
	$r_{ph}=  (\eta_c/\eta_0)^4r_{coast}$.

	Now, suppose that this outflow passes through a collimation shock located at some radius $r_s=10^{11}r_{s11}$ cm. 
	If $r_s\gg R_0/\Gamma_0\theta_0$, which is always the case once the jet breaks out from the envelope,
	the Lorentz factor of the fluid just upstream of the collimation shock satisfies $\Gamma \gg \theta_0^{-1}$.
	Upon shock crossing, the outflow decelerates until reaching the Lorentz factor
	$\Gamma_s\approx\theta_0^{-1}$ just behind the shock \citep{Bromberg2011b}\footnote{One way to
		derive this result is by employing the jump conditions of a relativistic, oblique shock \citep[e.g.,][]{henriksen1988}, that can be solved to yield the postshock 
		Lorentz factor, $\Gamma_s=3/\sqrt{8}\sin\psi$, and the deflection angle of streamlines across the shock, $\sin\delta=\cos(2\psi)/(1+8\cos^2\psi)^{1/2}$, in terms of the incidence angle $\psi$ of the unshocked fluid, measured with respect to the shock surface.  In the small angle approximation
		($\sin\delta  \ll1$) the latter relations reduce to $\Gamma_s \simeq (\sqrt{2}\sin\delta)^{-1}$. Since the shocked flow channel is cylindrical, $\delta=\theta_0$
		to a good approximation, and $\Gamma_s \simeq \theta_0^{-1}$.}.
	As long as the collimation shock is located inside the dense medium where the external pressure is high (as mostly happens in long GRBs), then the jet propagates above the collimation shock roughly cylindrically, with a constant cross-sectional radius, and thus constant Lorentz factor and temperature, until it emerges from the high pressure region. As soon as
		the cocoon's pressure declines the shocked fluid starts accelerating again and the jet becomes conical again, with the focus of the radial streamlines shifted from the injection point to a new location. We denote the radius at which the conical acceleration starts, as measured from the shifted focal point as $R_i$  (see Figure \ref{fig:sketch}). Since the opening angle of the re-accelerating flow is roughly $\theta_0$ and the propagation from the collimation shock to the re-acceleration point is roughly cylindrical, $R_i \approx r_s$.
	
	For illustration, suppose that the re-accelerating flow is conical with the same opening angle as the initial one, $\theta_0$ and that $R_i=r_s$. 
	Denote by $\eta_{s}$ the load of a fluid element behind the collimation shock. 
	If no additional baryon loading occurs during the collimation process, then $\eta_s=\eta_0$.  However, as we find below, mixing at the 
	collimation zone can increase the load, so that $\eta_s$ can vary among  different fluid elements, but must satisfy $\eta_s \le \eta_0$. 
	The new coasting radius will be located 
	at $r_{s,coast}\sim\eta_s r_s/\Gamma_s = \eta_s r_s \theta_0$.
	The photospheric radius of the re-accelerating fluid element, $r_{ph}$, will coincide with the new coasting radius 
	if $\eta_s = \eta_{s,c}$, where
	\begin{equation}
	\eta_{s,c}=\left(\frac{\sigma_T L_{jiso}\Gamma_s}{4\pi m_pc^3 r_s}\right)^{1/4} 
	=180 ~L_{iso,52}^{1/4}(r_{s11}\theta_{-1})^{-1/4}
	\label{eq:eta_sc}
	\end{equation}
	is obtained upon replacing $R_0$ and $\Gamma_0$ in Eq. (\ref{eq:eta_c-def1}) with $r_s$ and $\Gamma_s$, respectively.
	High radiative efficiency, roughly $1-r_{ph}/r_{s,coast}$, is anticipated  if $\eta_s > \eta_{s,c}$.   
	When $\eta_s<\eta_{s,c}$ the photosphere is located in the coasting zone, at $r_{ph}\approx (\eta_{s,c}/\eta_s)^4r_{s,coast}$,
	and the efficiency (prior to the onset of shell's spreading) is roughly given by $(\eta_s/\eta_{s,c})^{8/3}$ for $(\eta_s/\eta_{s,c})^{8/3}\ll1$ \citep{levinson2012}.  
	Hence, for a typical GRB with $L_{iso,52}\sim 1$, an efficient photospheric emission is expected unless	$\eta_s <100$, as will be confirmed by detailed 3D simulations below. 
	
	\subsection{Limits on observed temperature}
	\label{sec:temp}
	The temperature behind the collimation shock depends on the photon production rate in the immediate downstream.
	In terms of the photon-to baryon density ratio, $\tilde{n}=n_\gamma/n$, the enthalpy per baryon 
	can be expressed as $h=1+4p/nm_pc^2\approx 4\tilde{n}kT/m_pc^2 \gg1$, where the relation $p=n_\gamma kT$ has
	been employed and $T$ denotes the comoving temperature. Using $\eta_0=h\Gamma$, the observed temperature can be written as 
	$kT_{obs}=\Gamma kT=m_pc^2\eta_0/4\tilde{n}$. Note that in the absence of photon generation, mixing will not
	alter this relation, since $\tilde{n}$ will change by exactly a factor of $\eta_s/\eta_0$, and since $\tilde{n}$ 
	is conserved, the observed temperature behind the shock should equal the temperature at the origin, 
	specifically:
	\begin{equation}
	kT_{obs}=\Gamma_skT_s=\Gamma_0kT_0 \approx 800~L_{iso,52}^{1/4}R_7^{-1/2}\Gamma_0^{1/2} \quad {\rm keV}.
	\label{eq:T_obs}
	\end{equation}
	Photon generation behind the shock will give rise to a reduction in the observed temperature by
	a factor of $1+\Delta\tilde{n}/\tilde{n}$, where $\Delta\tilde{n}=\Delta n_{\gamma s}/n_s$ denotes
	the ratio of newly generated photons to baryons downstream.  In any case, the downstream temperature cannot 
	be lower than the black body limit,  $kT_{BB}= 10 (L_{iso,52}/\Gamma_s^2r_{s11}^2)^{1/4}$ keV, or
	transformed to the observer frame, $kT_{obs}>\Gamma_skT_{BB}\approx 30 (L_{iso,52}/\theta_{-1}^2r^2_{s11})^{1/4}$ keV.

To estimate  $\Delta\tilde{n}$ note that the relative number of newly generated photons behind the shock 
is given by $\Delta n_{\gamma}\simeq \dot{n}_{ff} t^\prime_s$, where $t^\prime_s=l/\Gamma_sc$ 
is the proper flow time of the shocked plasma as it crosses the distance $l$ between the collimation shock and re-acceleration zone,
$\dot{n}_{ff}\simeq \alpha_e\sigma_T c (1+x_\pm)^2n_s^2(kT_s/m_ec^2)^{-1/2}\Lambda_{ff}$ 
is the approximate free-free emission rate (e.g., \citealt{levinson2012}), $n_s$ and $T_s$ are the proper baryon density and temperature behind the collimation shock, respectively, $x_\pm$ is the pair-to-baryon ratio, $ \Lambda_{ff} $ is a logarithmic factor that accounts for upscatter of low-energy free-free photons by inverse-Compton. For illustration we adopt $\Lambda_{ff}\simeq10$. In terms of the pair unloaded optical depth behind the collimation shock, $\tau=\sigma_T n_s l/\Gamma_s$, 
one then obtains $\Delta n_\gamma/n_s\simeq 0.2(1+x_\pm)^2 \tau  (kT_s/m_ec^2)^{-1/2}\approx 0.6(1+x_\pm)^2\tau$ for the normalization adopted in  
Eq. (\ref{eq:T_obs}).  At rest-frame temperatures above $50$ keV roughly the pair density becomes large, $x_\pm>>1$. 
Compared with the density ratio at the origin,  $\tilde{n}=m_pc^2\eta_0/4kT_0\Gamma_0\approx 5\times10^{4}(\eta_0/\eta_{s,c})(r_{s11}\theta_{-1})^{-1/4} (R_7/\Gamma_0)^{1/2}$, it is seen that  even a modest $\tau$ is sufficient to increase the photon-to-baryon ratio, $n_\gamma/n_s$, well above that produced at the outflow injection point.  A rough estimate of the optical depth behind the shock yields
\begin{equation}\label{eq:tau}
		\tau\sim \eta_{s,c}^4 l/\eta_s\Gamma^3_s r_s \sim 10^{4}(\eta_{s,c}/\eta_s)L_{iso,52}^{3/4}r_{s11}^{-3/4}\theta_{-1}^{9/4} (l/r_s)~.
		\end{equation}
For typical parameters $l$ and $r_s$ are similar to within an order of magnitude so $l/r_s \sim 1$. 
For the fiducial simulation  $A$ in table \ref{tab_models_comparison} we find $\tau\simeq3\times10^5$, somewhat above this value.
Hence it is expected that as long as the proper temperature behind the collimation shock is larger than 50keV, $ x_\pm \gg 1 $ and photon production is efficient enough to reduce the temperature. Thus, pair creation regulates the rest-frame temperature behind the shock not to exceed 50keV.
The observed temperature would depend on the opening angle of jet: $kT_{obs}=\Gamma_s kT \simeq 50 \theta_0^{-1} $ keV.
For case A we find $\Gamma_s\simeq4$, implying $kT_{obs}\sim 200 $ keV.  Note 
that $L_{jiso}\propto \theta_0^{-2}\propto \nu_{p}^2$, where $h\nu_p=kT_{ obs}$ is the photon energy at the spectral peak.
Interestingly, this is consistent with the Amati relation (but likely altered by viewing angle effects, see section \ref{sec:summary}).

	\section{Numerical Simulations}
	\label{sec:simulations}
	
	In order to test our analytic result and better quantify the properties of the outflow near the photosphere, we performed a set of 3D RHD simulations for a range of the model parameters (the jet power $L_j$, initial opening angle $\theta_0$ and 
	initial load $\eta_0=h_0\Gamma_0$).  The output of these 
	simulations is used in $ \S $ \ref{sec:efficiency} to compute the radiative efficiency 
	along different sight lines.  All simulations have been carried out using 
	the PLUTO v4.2 code \citep{Mignone2007}, where we applied an equation of state of a relativistic 
	gas with an adiabatic index of 4/3.  
	In this section we briefly describe the results of the simulations, 
	focusing on key features that are relevant to the analysis presented in $ \S $ \ref{sec:efficiency}.  
	A detailed account of the full setups of the 3D Cartesian grids, convergence tests and numerical results will be given in Gottlieb et al., in prep.
	
	\subsection{Numerical Setup \& Models}
	\label{sec:models}

	We consider both lGRBs (models $ \A - \E$) and sGRBs (model $ \F $) 
	with the parameters listed in Table $ \ref{tab_models_comparison} $. 
	In all lGRB simulations we invoke a static, non-rotating progenitor star of radius $ R_\star = 10^{11}\cm $, mass $ M_\star = 10\msun $ and density profile $ \rho(r) \propto r^{-2}x^3 $, where $ x \equiv (R_\star-r)/R_\star $.   The jet in models $A-E$ 
	is injected from a radius $R_0=10^{-2}R_\star$ with an initial Lorentz factor $ \Gamma_0 $ into a cylindrical nozzle, beyond
	which it quickly opens to a final angle $ \theta_0 = 1/f\Gamma_0 $, with $ f = 1.4 $ \citep{Mizuta2013,Harrison2017}.
	In all cases the simulation ends long after the jet breaks out of the 
	star (see table \ref{tab_models_comparison}), typically when the jet's head reaches
	a radius of 10-15 stellar radii, but before most fluid elements reach the photosphere.  
	As will be explained below, we find this to be sufficient, since 
	by this time most of the outflow has begun the second homologous expansion phase and its future evolution can
	be extrapolated analytically. Reaching the photosphere in a 3D simulation with a sufficient resolution requires 
	tremendous resources or impractically long runs that we find unnecessary.

	Model $F$ corresponds to a sGRB jet that emerges from a double neutron star merger. It includes three components:
	(a) A relativistic jet launched from the origin with a delay in respect to the merger time.
	(b) A non-relativistic ($ v_c < 0.2c $) cold core ejecta with mass $ M_c = 0.05 \msun $ and density profile $ \rho(r)\propto r^{-2} $. (c) A mildly-relativistic cold tail ejecta as indicated in previous works \citep{Hotokezaka2012,Hotokezaka2018,Kyutoku2012,Bauswein2013,Beloborodov2018} with mass $ M_t \approx 0.05 M_c $ and density profile $ \rho(r)\propto r^{-14} $.
	Both components of the ejecta are homologous.

	\begin{table}
		\setlength{\tabcolsep}{6.8pt}
		\centering
		\begin{tabular}{ | l | c  c c  c  c  c | }
			\hline
			lGRB  & $ L_j  $ & $L_{j,iso}$ &$ \theta_0 $ & $ \eta_0 $ & $ t_b  $ & $ t_e $ \\
			model& $ [\rm{erg~s^{-1}}] $ &$ [\rm{erg~s^{-1}}]$ &  & & $ [\s] $ & $ [\s] $ \\ \hline
			$ \A $ & $ 10^{50} $ & $10^{52}$& $ 0.14 $ & $ 500 $ & $ 20 $ & $ 68 $ \\
			$ \B $ & $ 5 \times 10^{50} $ & $5 \times 10^{52}$ &$ 0.14 $ & $ 500 $ & $ 8 $ & $ 32 $ \\
			$ \C $ & $ 10^{50} $ & $3.5 \times 10^{51}$ &$ 0.24 $ & $ 300 $ & $ 27 $ & $ 57 $ \\
			$ \D $ & $ 10^{50} $ & $3.5 \times 10^{51}$ &$ 0.24 $ & $ 500 $ & $ 28 $ & $ 70 $ \\
			$ \E $ & $ 10^{50} $ & $10^{52}$ &$ 0.14 $ & $ 100 $ & $ 13 $ & $ 43 $ \\
			\hline
			sGRB  & $ L_j  $ & $L_{j,iso}$ & $ \theta_0 $ & $\eta_0 $ & $ t_d; t_b $ & $ t_e $ \\
			model& $ [\rm{erg~s^{-1}}] $ &$ [\rm{erg~s^{-1}}]$ & &  & $ [\s] $ & $ [\s] $ \\ \hline
			$ \F $ & $ 3 \times 10^{49} $ & $ 3 \times 10^{51} $& $ 0.14 $ & $ 500 $ & $ 0.6; 1.4 $ & $ 3.0 $ \\ \hline
			
		\end{tabular}
		\hfill\break
		
		\caption{The simulations configurations. $ L_j $ is the total jet luminosity, $ \theta_0 $ is the jet opening angle at injection in radians, $ \eta_0 \equiv \Gamma_0h_0 $ (assuming relativistic motion) is the terminal four velocity of the jet which is defined by the initial Lorentz factor $ \Gamma_0 $ and the initial specific enthalpy $ h_0 $, $ t_d, t_b $ are the delay time and the breakout time (from the star in lGRBs or the core ejecta in sGRBs), respectively.
		In the sGRB model $ \F $, $ t_d $ and $ t_b $ are measured in respect to the time of the merger.
		In the lGRB models, the engine time $ t_e $ is also the total time of the simulation, while in the sGRB model the total time of the simulation is $ t_d + t_e = 3.6 $s.}\label{tab_models_comparison}
	\end{table}

	\subsection{Simulation Results \& Calculation Method}
	\label{sec:simulation_results}

	The structure of the outflow (shown in Figure \ref{fig:sketch} for lGRBs) in all runs features a strong (primary) collimation shock, followed by
	one or more weaker shocks just above the primary shock.  Beneath the primary shock (that is, between
	the injection point and the shock) the 
	outflow expands conically with an opening angle $\theta_0$, while above it the weak shocks keep it roughly cylindrical.  The collimation shocks propagate slowly 
	outwards as the system evolves, however, in all lGRB simulations presented here the primary shock remains inside the star, reaching a radius $ r_s\sim 0.5R_\star $, by the end of the simulation.
	While for very powerful jets the primary shock may break out \citep{Lazzati2009}, we find that for typical lGRBs this is not the case. This 
	implies that the shock has a large optical depth during the prompt emission phase (see \S \ref{sec:previous} for further discussion), which has important implications for the emitted spectrum that will be discussed later on. 
	The weak collimation shock extends, at late times, just above the star, up to a radius of about $2R_\star$ (marked as weak collimation zone in Fig. \ref{fig:radii}), further delaying the onset of the second homologous expansion phase.
	Only above this radius the outflow is expected to start expanding conically again. 
	Mixing at the collimation zone (see \S \ref{sec:mixing} for details) leads to a substantial stratification of the flow owing to additional baryon loading that depends on angles.  We carefully checked the trajectories of different 
	fluid elements\footnote{PLUTO in an Eulearian code, so that tracing the elements is done under the assumption that the temporal sampling is high enough so that the velocity of each cell does not significantly change between two snapshots.} and verified that, indeed, above the weak collimation zone 
	they all accelerate along radial streamlines, and that their Lorentz factor 
	increases linearly with radius, albeit its value varies among the different elements.  
	The additional loading at the collimation zone is sporadic and varies with time; it is at its maximum shortly after 
	jet breakout, but then our simulations show that it gradually declines (on average) as the system evolves, ultimately reaching a fixed average value 
	(about a factor of 2 to 3 smaller than the initial one).  
	As a result, fluid elements that emit at later times have higher radiative efficiency (see Fig. \ref{fig:radii}). 
	
	As stated above, most fluid elements do not reach the photosphere by the end of the simulations. However, 
	owing to their radial expansion above the weak collimation zone it is possible to extrapolate their 
	density, internal energy and Lorentz factor to radii outside the simulation box, thereby finding 
	the location of their photosphere and the specific enthalpy $h$ at the photosphere.  Our procedure is as follows: 
	First, we examine the acceleration of each fluid element in the simulation and check whether it is in the free acceleration zone, namely maintains $ \Gamma \propto r $. Once we verify that the element has indeed reached 
	the free expansion zone
	we use the analytic solution of an adiabatic fireball to determine its further evolution in time.  
	To be concrete, for each simulation time, $ \ts $, and for each element at some position $ {\bf r}_i(t_{sim})=(r_i,\theta_i,\phi_i) $ and velocity $ v =c$ we advance the parameters
	of this element along its radial trajectory, in the direction $\hat{\Omega}=\dot{{\bf r}}_i/|\dot{\bf r}_i|$, 
	to a radius ${\bf r}={\bf r}_i+\hat{\Omega}[t-t_{sim}]c$ at time $ t $ according to:
	\begin{eqnarray}
	\Gamma(t,{\bf r})=\Gamma(t_{sim},{\bf r}_i) \tilde{\Gamma}(s/R_i),\nonumber\\
	h(t,{\bf r})=h(t_{sim},{\bf r}_i)\tilde{h}(s/R_i),\\
	n(t,{\bf r})=n(t_{sim},{\bf r}_i) \tilde{n}(s/R_i).\nonumber
	\end{eqnarray}
	where  $\tilde{\Gamma}$, $\tilde{h}$, $\tilde{n}$ are universal profiles obtained from the integration of the adiabatic fireball equations, subject to the boundary conditions $\tilde{\Gamma}(1)=\tilde{h}(1)=\tilde{n}(1)=1$ (see appendix \ref{sec:numerical_calculation} for details), $s=[t-t_{sim}]c$ is the distance traversed by that fluid element along the direction $\hat{\Omega}$ over time $t-t_{sim}$, and $R_i=r_i\sin\theta_i/\sqrt{1-\Omega_z^2}$ is the effective injection radius.
	We check this extrapolation by advancing fluid elements from different simulation times, $ \ts_1 $ and $ \ts_2 $, to the same future time $ t \gg \ts_1, \ts_2 $, confirming that the same result at time $ t $ is indeed obtained.
	The optical depth to infinity is then given by $\tau(t,{\bf r})=\int_{\bf r}^\infty\sigma_T n(t,r)\Gamma(t,r)(1-v(t,r)/c)dr$.
	The photospheric radius $ r_{ph} $ of each element is defined as the location at which $ \tau = 1 $, whereby the specific enthalpy at the photosphere at time $t=t_{sim}+|{\bf r}_{ph}-{\bf r}_i|/c$ is given by $h_{ph}(t,\hat{\Omega})=h(t,{\bf r}_{ph})$.
	Finally, for each element at each angle and time we transform to the observer's frame to obtain the radiative efficiency as a function of the observer's time $ \tobs \equiv t - {\bf r}_{ph}/c$, where $ t $ is the time measured in the star frame, and we assume that beaming allows each observer to detect only emission that is emitted at the observer's viewing angle.

	The medium to which sGRB jets are injected is less dense in comparison to lGRBs. Consequently, the evolution of such systems is slightly different as they show jets which propagate faster, are subject to less mixing and produce less massive cocoons. This in turn results in less loading and lower optical depth in these systems.
	We performed numerical calculations of the radiative efficiency for one sGRB model as well. In this simulation the jet is injected into the system with a delay of 0.6s relative to the time of the merger. The jet propagates in an expanding cold ejecta with two components: a massive non-relativistic core ejecta and a dilute mildly-relativistic tail ejecta. The jet breaks out from the core (tail) ejecta $ \sim 1\s \ (2.5\s) $ after its launch at $ z \approx 8\times 10^9\cm \ (6.5\times 10^{10}\cm) $. Then the jet accelerates and at $ \sim 2 \times 10^{11}\cm $ reaches the photosphere to emit radiation.
	
	One important difference that we find between lGRB and sGRB models is the location and dynamics of the collimation shock. While in lGRBs the collimation shock remains inside the star for times longer than a typical burst duration 
	(with the exception of especially energetic jets, see further discussion in $ \S $ \ref{sec:previous}), 
	in sGRB models the collimation shock seems to break out rather quickly and propagate outwards in most cases (see our model F, and additional examples in \citealt{Gottlieb2018a,Gottlieb2018b}).
	If the collimation shock reaches a region of moderate optical depth during the emission phase, it can alter the emitted spectrum significantly \citep{ito2018b}.
	In our simulation F the collimation shock emerges from the core ejecta with a velocity of $ \sim 0.5 c $ at $ z = 2.7\times 10^{10}\cm $, 3s after the merger.

	\subsection{Mixing}
	\label{sec:mixing}
	
	\begin{figure*}
		\centering
		\includegraphics[scale=0.23]{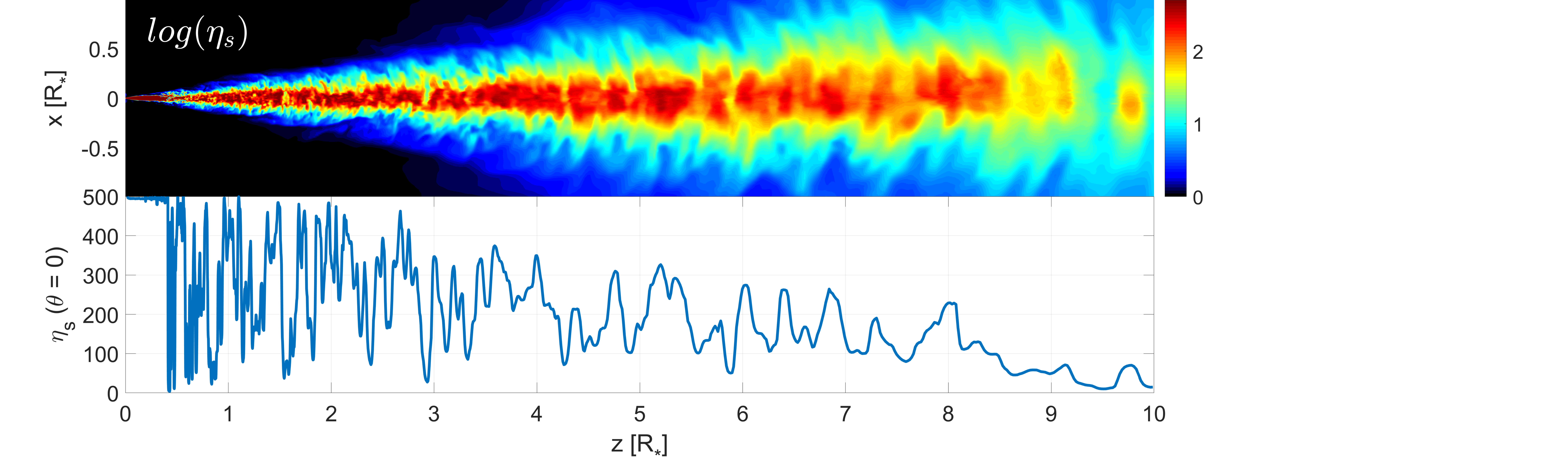}
		\caption[]{The spatial distribution of the load parameter $ \eta_s $ in model $ \A $ at Lab time $t = 55\s$, at which the jet's
			head has reached $ 10 R_\star$.  The upper panel shows a full map of $ log(\eta_s) $, 
			and the lower panel presents the profile of $ \eta_s $ along the $z$-axis (at $\theta=0$).
		}
		\label{fig:map}
	\end{figure*}
	
	We find that the jet structure and evolution are greatly affected by the jet's mixing. The degree of mixing is also essential for the prompt $ \gamma $-ray signal as it dictates the efficiency of the photospheric emission. The curved paths of jet elements along the collimation shock include lateral motion from and into the cocoon. This motion induces the formation of Rayleigh-Taylor \citep{Rayleigh1882,Taylor1950} fingers along the jet-cocoon interface. These instabilities grow over time, increasing the degree of mixing between jet and cocoon matter, thereby enhancing the baryon loading in the jet \citep{Matsumoto2013,Matsumoto2017,Toma2017}. Once the jet-cocoon structure breaks out from the dense environment, it expands and accelerates, while each element keeps its loading (i.e., mixing ends as the re-acceleration begins). Following jet elements that were launched at different times we find that in our models the mixing is less severe in elements that are launched at later times. Namely, $ \eta_s $ is typically smaller for elements that are at the jet front, and is larger for elements that are launched after the jet breaks out, albeit with large fluctuations between different elements. The resulting emission is expected to show a temporal evolution from low to high efficiency. We shall report a detailed analysis of the mixing in a future paper.
	
	The top panel in Figure \ref{fig:map} depicts a map of $ \log(\eta_s) $ of model $ \A $ when the jet head reaches $ 10R_\star $. The red (cyan) region reflects the jet (jet-cocoon interface).
	The second top panel shows $ log(\eta_s) $ along the injection axis.
	It is prominent that the jet is very inhomogeneous due to the intense mixing it underwent inside the star.
	The high variability begins just behind the collimation shock at $ \sim 0.5R_\star $ and continues all along the jet. Note that the decreasing variability in the figure is due to logarithmic cells distribution and not to the physics.
	The velocity difference between adjacent elements will eventually result in internal shocks.
    %	We estimate that the net energy dissipating in these internal shocks amounts to no more than a few percents of the total jet energy. 
	Nonetheless, these shocks can have a profound effect on the emitted spectrum \citep[e.g.,][]{keren2014,ito2018b}. Such an analysis is beyond the scope of this paper, and will be discussed in a future work.

	\section{The Efficiency of Photospheric emission}
	\label{sec:efficiency}
	
	Since the internal energy of the flow is dominated by radiation, the power emitted into a solid angle $d\Omega$ in the direction $\hat{\Omega}$ 
	at time $t$, can be expressed as 
	\begin{equation}
	dL_\gamma(t,\hat{\Omega})= n_{ph}(t,\hat{\Omega})m_pc^3[h_{ph}(t,\hat{\Omega})-1]\Gamma^2_{ph}(t,\hat{\Omega}) r^2_{ph}(t,\hat\Omega)d\Omega
	\end{equation}
	where $r_{ph}(t,\hat{\Omega})$ is the photospheric radius of fluid moving in the direction $\hat{\Omega}$ at time $t$, 
	and $n_{ph}(t,\hat{\Omega})$, $h_{ph}(t,\hat{\Omega})$, $\Gamma_{ph}(t,\hat{\Omega}) $
	denote the proper density, enthalpy per baryon and Lorentz factor at the photosphere at time $t$.   This luminosity constitutes a fraction
	\begin{equation}
	\epsilon(\hat{\Omega},t)=\frac{h_{ph}(t,\hat{\Omega})-1}{h_{ph}(t,\hat{\Omega})}
	\label{eq:efficiency_def}
	\end{equation}
	of the corresponding jet power, $dL_j(t,\hat{\Omega})$, that we henceforth define as the radiative efficiency (and denote for short by $\epsilon$).
	In practice the efficiency can be higher if some form of dissipation (not included in our model) deposits energy just beneath the photosphere,
	so formally the quantity define in Eq. (\ref{eq:efficiency_def}) is the minimum efficiency.  
	%%%%
	
	The computation of $\epsilon(\hat{\Omega},t)$ for each fluid element follows directly from the analytic 
	extrapolation of the simulation output to the photosphere, as explained in detail in the preceding section.   
	In the following we describe, in turn, the results obtained for lGRBs (models $ \A-\E $) and sGRBs (model $ \F $).
	We present only results obtained in the $\phi=0$ plane, which we find to be representative of all other sight lines.

	\subsection{lGRBs}
\begin{figure*}
		\centering
		\includegraphics[scale=0.23]{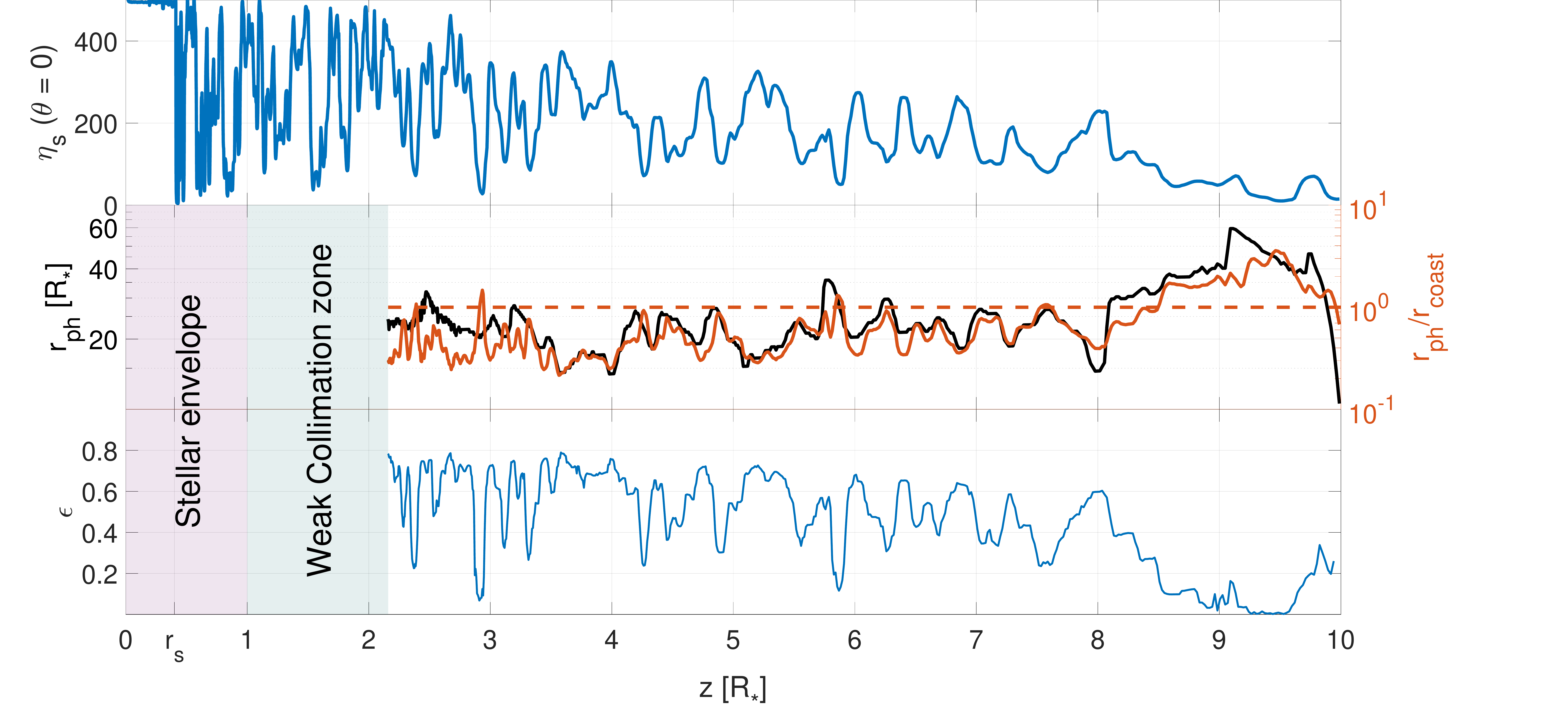}
		\caption[]{Results for model $ \A $ at Lab time $ t=55\s$, at which the jet's head has 
			reached a radius of $ 10R_\star $,  with the plug omitted.
			The middle panel displays a mapping between the extrapolated photospheric radius of fluid elements on-axis and their location at time $ t=55\s$ (black line), and the bottom panel shows the corresponding radiative efficiency. The red line in the middle panel gives the ratio of the photospheric radius and the coasting radius.  The top panel shows the same $ \eta_s $ profile as in the lower panel of Figure \ref{fig:map}, and is presented here to elucidate the relation between mixing and efficiency.
			The pinkish and greenish areas on the left mark the stellar interior and the weak collimation zone, respectively. The radius of the primary collimation shock, $r_s$, is also indicated.
			Free expansion recommences above the weak collimation zone, where our extrapolation scheme applies. 
		}
		\label{fig:radii}
	\end{figure*}

\begin{figure*}
		\centering
		\includegraphics[scale=0.2]{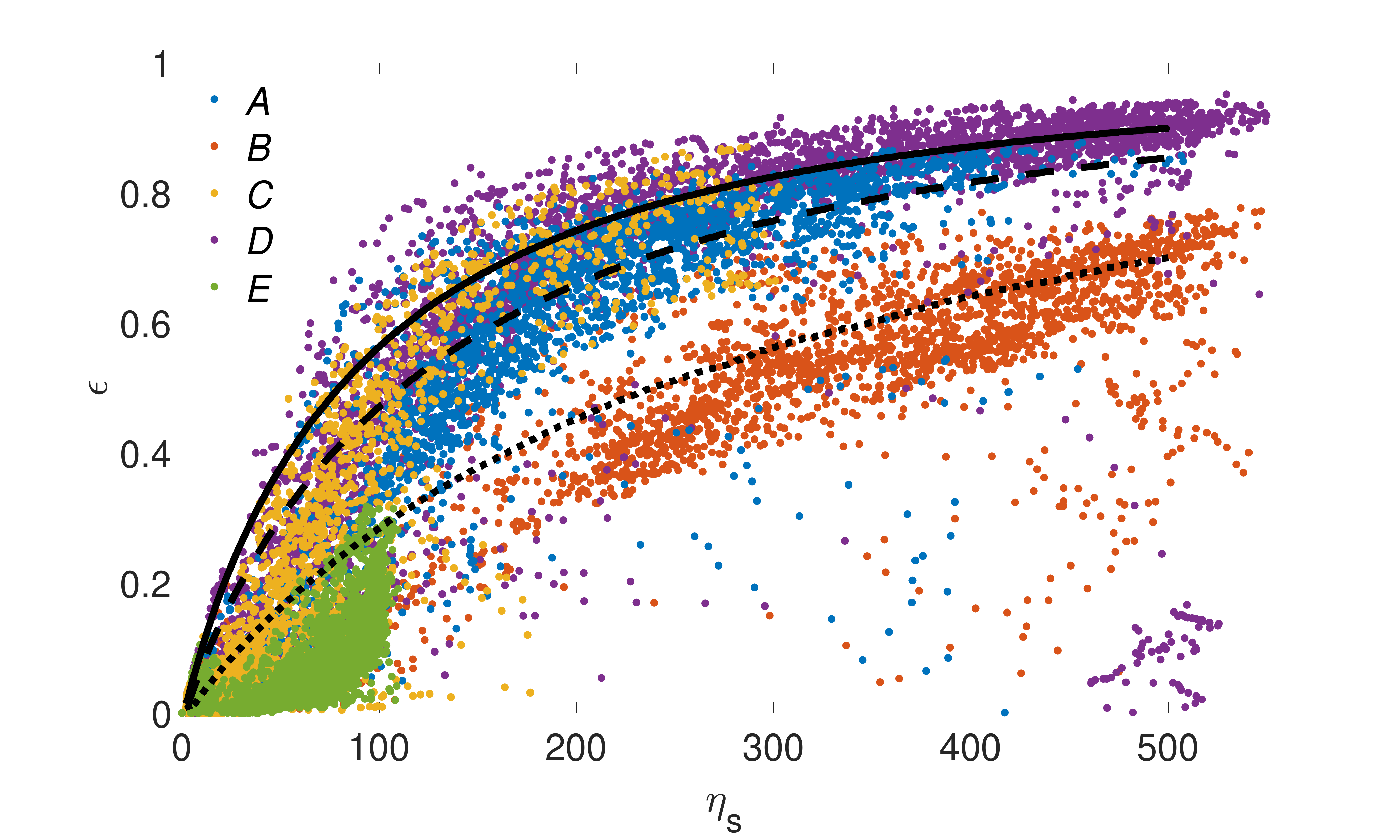}
		\includegraphics[scale=0.2]{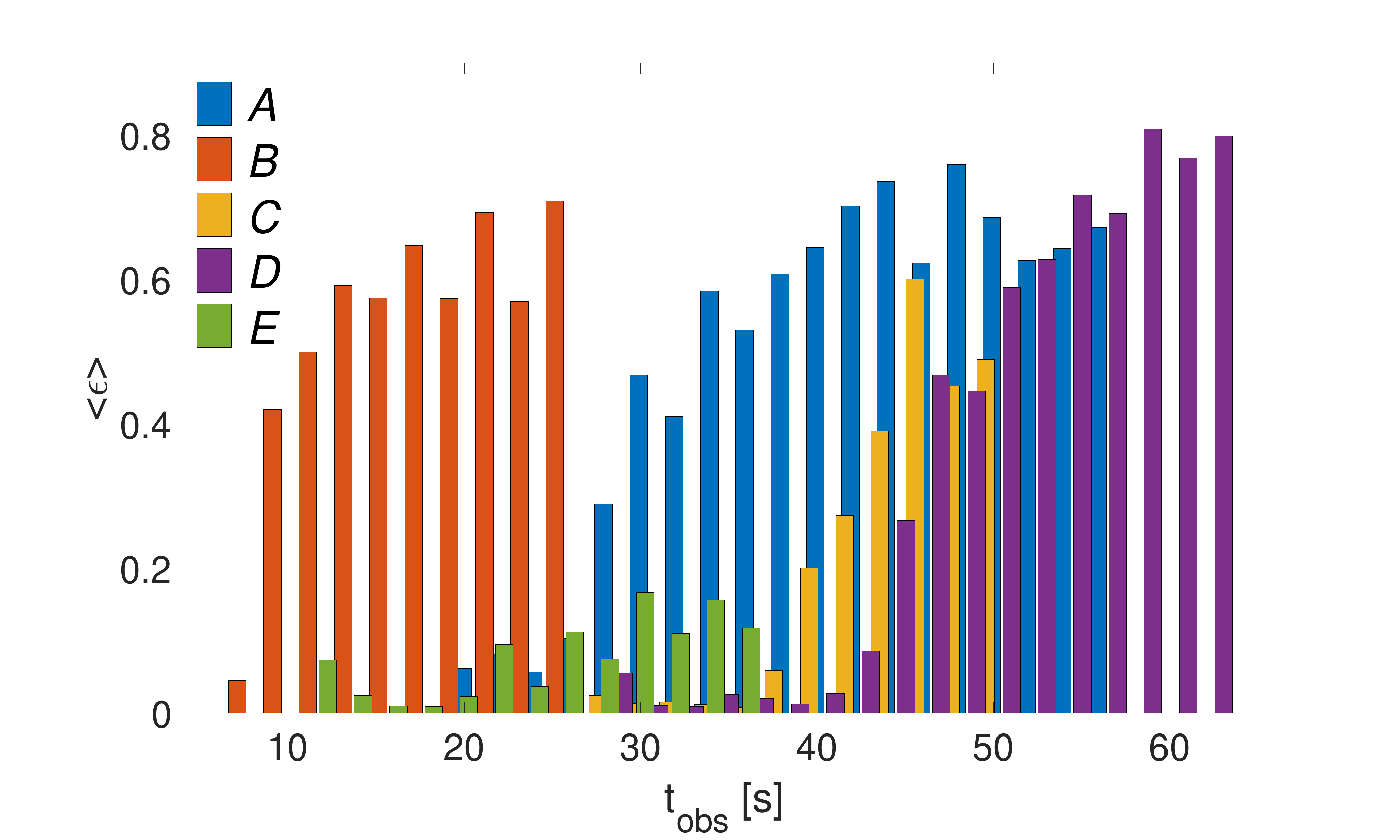}
		\caption[]{Left: The dependence of the radiative efficiency on the load parameter $\eta_s$ obtained from the simulations, for fluid elements along the jet axis.
			The variation in $\eta_s$ between different fluid elements in each model is caused by mixing at the collimation throat.  The black lines delineate the analytic result obtained from the integration of the adiabatic fireball equations (appendix \ref{sec:numerical_calculation}) for the parameters
			in table \ref{tab_models_comparison} (the solid line corresponds to models $ \C$ and $ \D $, the dashed line to models $ \A $ and $ \E $, and the dotted line to model $ \B $; see text for further details).
			Right: The efficiency as a function of observer time, presented in bins of two seconds for clarity.  The observer time is measured with respect to the explosion (i.e., as the jet launch starts) to show the full delay in the
				onset of emission.  This delay implies a minimum engine operation time to reach high efficiency. 
		}
		\label{fig:lGRBs_models}
	\end{figure*}

	We first consider the efficiency along the jet's axis, where it is expected to be the highest.
	Figure \ref{fig:radii}  depicts the  extrapolated photospheric radius (middle panel) and efficiency 
	(bottom panel) of fluid elements in model $ \A $ as a function of their location at a particular simulation time ($t=55$s in this case);
	note that fluid elements located at larger distances $z$ crossed the collimation 
	shock earlier than elements located at smaller distances. 
	Also shown in the middle panel (red line) is the ratio of the photospheric radius and coasting radius, which is directly related to the efficiency. 
	The top panel exhibits the corresponding load of the fluid elements (it  is identical to the bottom panel in Figure \ref{fig:map}) and is 
	presented here to conveniently show the relationship between mixing and efficiency. 	
	As seen, the efficiency features large fluctuations around a rather high average value, about $0.6$, 
	except for fluid elements near the jet's head  (around $z=9 R_\star$).   The particularly low efficiency 
	of those elements reflects the relatively large loading by mixing near the collimation shock at early time, during jet breakout from the star (see discussion in \S \ref{sec:mixing}).   
	Over time  mixing diminishes and the efficiency increases (see Figure \ref{fig:lGRBs_models} for temporal evolution). 
	%%%

	%%%%%%%%%%%%%%%
	
	A comparison of the models is presented in  Figure \ref{fig:lGRBs_models}.
	The left panel depicts the relation between the radiative efficiency and the loading parameter $\eta_s$ for the set of on-axis fluid elements 
	in each model.   From the data in table \ref{tab_models_comparison} and Eq. (\ref{eq:eta_sc}), with $r_{s11}=2$ for models $A, C-E$ 
	and $r_{s11}=1$ for model $B$ (adopted from the simulations), we obtain $\eta_{s,c}\approx100$ for
	models $\C $ and $ \D $,  $\eta_{s,c}\approx 160 $ for models $ \A $ and $ \E $, and $\eta_{s,c}\approx 300$ for model $ \B $.  
	We find that the  shape of the $\epsilon-\eta_{s}$ curve for each model is in a good agreement with the analytic result (the black lines) 
	obtained from integration of the adiabatic fireball equations (appendix \ref{sec:numerical_calculation}) with the corresponding $\eta_{s,c}$ for each model.
	It is worth noting that, while the scatter around the analytic curve in each model is not so large, 
	some fluid elements show substantial deviation from it.   One particular example is the ``island"  around $ \epsilon \approx 0.2 $ and $ \eta_s \approx 500 $ in model $ \D $. The reason for the large deviation is that those fluid elements, which are located just beneath the head, encounter the opaque plug \citep{Zhang2003} (i.e., the material accumulated in front of the jet's head during its propagation inside the star) prior to reaching their own photosphere, which significantly suppresses their radiative efficiency despite their relatively large $\eta_s$. This is more apparent in model D by virtue of the larger jet opening angle that gives rise to a more massive plug.	
	However, this large suppression occurs early on, and as time goes by, the plug becomes more transparent and its effect diminishes. 

\begin{figure}
		\centering
		\includegraphics[scale=0.21]{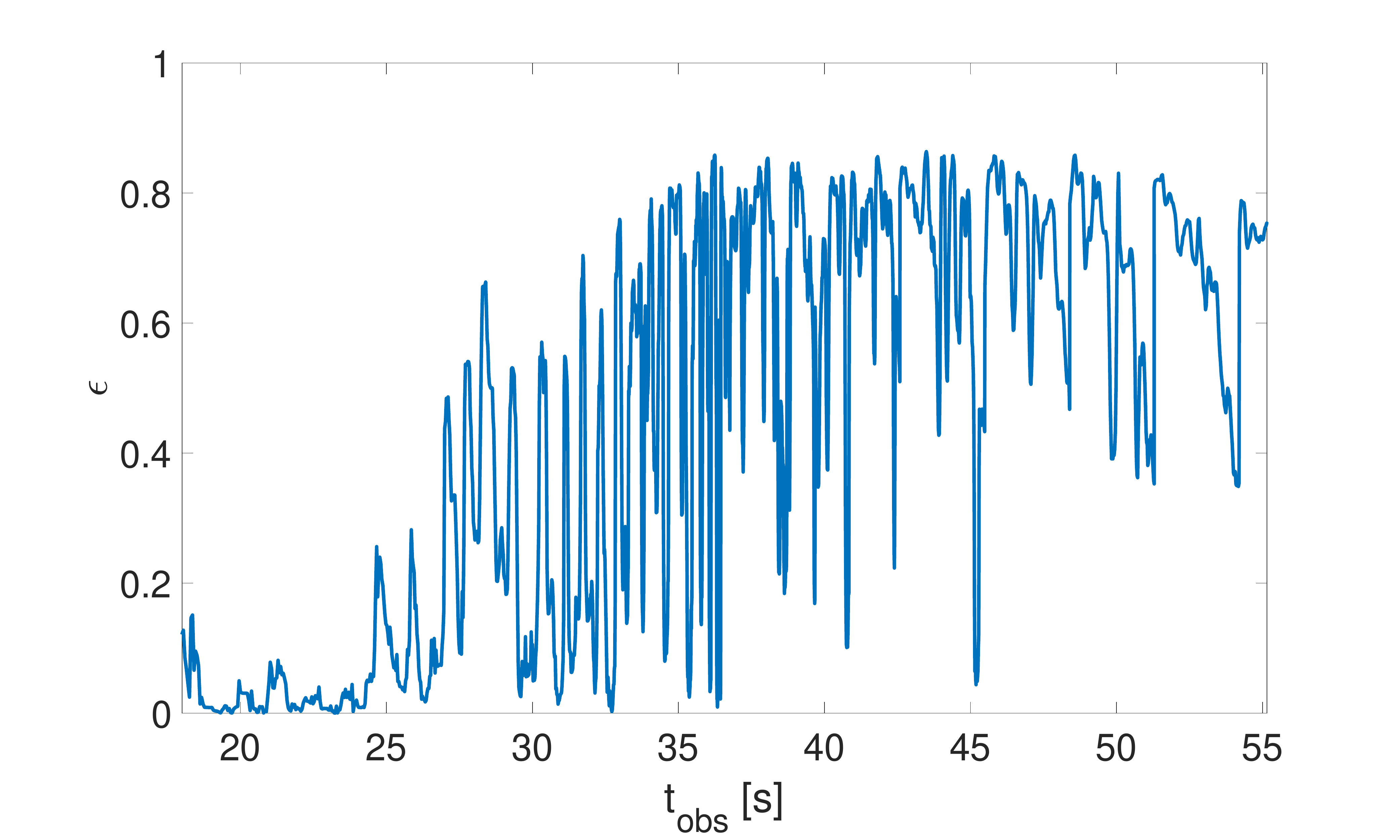}
		
		\caption[]{The radiative efficiency on-axis in model $ \A $ as a function of the observer's time.
		}
		\label{fig:variability}
	\end{figure}

	The right panel in Figure \ref{fig:lGRBs_models} shows the temporal evolution of the  efficiency in the observer's frame, averaged over 2s bins for clarity (the full unbinned plot of simulation $ \A $ in presented in Figure \ref{fig:variability} for illustration). Note that $ \tobs = 0 $ is the arrival time of a virtual photon emitted from the origin as the jet launching starts. This is not the actual arrival time of the first photon observed, which can be emitted only after the jet breaks out of the star, roughly with a delay of $t_b-R_\star/c$.	
	As seen, a gradual increase of $\epsilon$ commences when the first fluid elements start reaching the photosphere.    The delay in the onset of emission is dictated primarily by the jet breakout time (see table \ref{tab_models_comparison}) and, to a lesser extent, by mixing.   It defines the minimum engine operation time required to have an efficient photospheric emission. 
	Note that in all models except for model $E$ the initial load is smaller than the critical load at the collimation shock, that is $\eta_0 > \eta_{s,c}$. 
	However, even in model $E$ the efficiency can reach as high as $20\%$.  

	The results described above indicate that for a wide range of model parameters strong photospheric emission cannot be avoided when observing the jet nearly on-axis, unless loading at the origin is so large that it limits the asymptotic Lorentz factor to values well below $100$,  in tension with compactness arguments in certain cases.
	
	We turn now to analyze the efficiency along off-axis sight lines ($\theta>0$)\footnote{The angles are henceforth measure with respect to the shifted origin, obtained by interpolation of the radial streamlines above the weak collimation shock.}.  
	Naively, it is anticipated to decline with angle beyond the jet core, owing to stratification of the flow by mixing. 	
	From the simulations we identify a re-opening core angle above the weak collimation shock of $ \theta_n \approx 0.3\theta_0 - 0.5 \theta_0 $, depending on $\theta_0$.
	Within this angle the average loading by mixing is roughly uniform across the jet.  Beyond this angle 
	lies the heavily  mixed cocoon material for which the specific enthalpy is smaller and the optical depth is larger. 
	We find that the photospheric radius of material flowing along the $ \theta_n $ direction is, on average, twice 
	as large as that of on-axis fluid, and  at $ 2\theta_n $ it is four times larger.
	
\begin{figure*}
		\centering
		\includegraphics[scale=0.2]{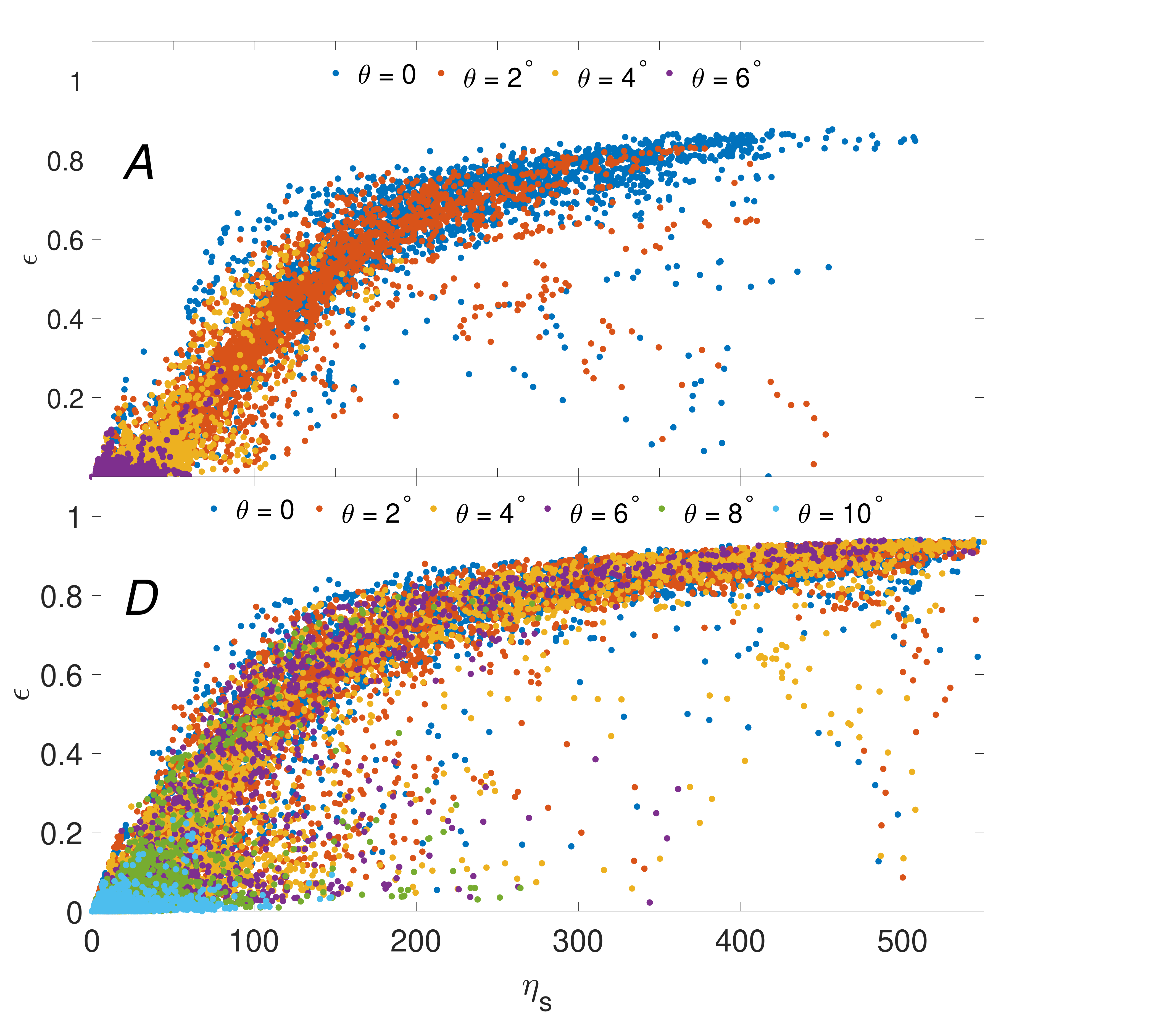}
		\includegraphics[scale=0.2]{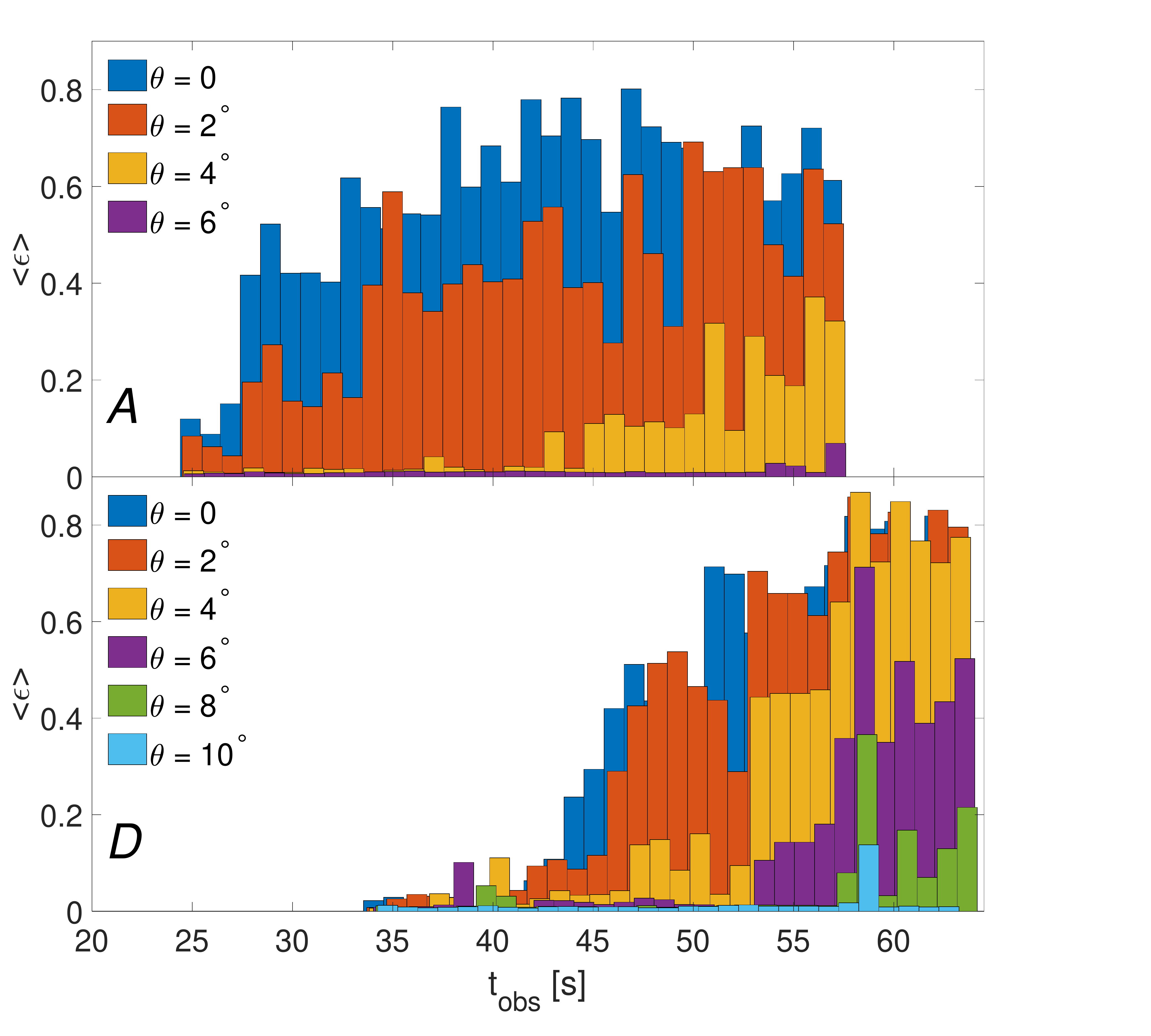}
		\caption[]{The efficiency at various angles for lGRB models $ \A $ (top) and $ \D $ (bottom).
			Left: The dependence of the efficiency on the load parameter $ \eta_s $.
			Right: The efficiency as a function of observer time, plotted in bins of one second for clarity.
			In model $ \A $ $ \theta_n \approx 3^\circ \approx 0.4\theta_0 $, whereas in model $ \D $ $ \theta_n \approx 5^\circ \approx 0.4\theta_0 $.
		}
		\label{fig:lGRBs_models_angles}
	\end{figure*}

	Figure \ref{fig:lGRBs_models_angles}  shows the radiative efficiency in models $A$ (upper panels) and  $ \D $ (lower panels) along different directions $\theta$.   The left panels indicate that the $\epsilon-\eta_s$ curves follow the analytic results, confirming that free
	expansion of streamlines occurs also at these larger angles.   The right panels exhibit the temporal evolution, and it is seen that 
	a rather large efficiency is expected up to a few degree, albeit with a larger delay compared with the core emission.
	A comparison of the two models verifies that there is a dependence of $\epsilon$ on the re-opening angle of the jet, $\theta_n$. Within this angle the efficiency is at its maximum and is roughly uniform.  At larger angles it drops sharply. One can also see that the characteristic time over which the efficiency reaches its maximum increases with increasing viewing angle. In our simulations the efficiency is still growing at large angles at the end of the simulation. Therefore, it is possible that for longer engine times higher efficiency will be obtained in those directions, and perhaps even at larger angles than the ones presented here.
	%We find that the efficiency can reach a few percents even at $\theta_0$.
	%	
	
	Ultimately the radiative efficiency should be manifested in the observed light curve of photospheric emission.	Interestingly, we find quite rapid temporal changes of the efficiency, with typical timescales of a few tens of ms, consistent with the characteristic variability timescales observed in lGRBs \citep{MacLachlan2012,Bhat2013}.
	We stress that our results may be limited by the resolution, and that higher resolution may reveal shorter variability times. This issue will be addressed in a follow up paper.
	
	\subsection{sGRBs}
	\label{sec:sgrb}
	
	Our sGRB configuration in model $ \F $ includes massive core ejecta and dilute tail ejecta. We find that the latter is not important for the jet evolution or to the location of the photosphere as it is dilute and optically thin for the photons emitted from the jet\footnote{Other velocity and density profiles of tail ejecta could affect the optical depth of photons. In particular, a shallower density profile would increase the optical depth of the tail ejecta, thereby pushing the photosphere further out, reducing the efficiency.}.
		Since the core ejecta is less dense than the stellar environment in lGRBs, the sGRB jet undergoes less mixing compared to lGRBs. This induces a lower optical depth, and consequently smaller $ \rph/r_{coast} $ values. Furthermore, the dilute core ejecta leads to a collimation shock that is very high, so that the jet is only marginally collimated (there is no weak collimation zone), thereby increasing the specific enthalpy. Both effects result in higher efficiency in sGRBs than lGRBs. At late times the collimation shock almost reaches the photosphere and may alter the spectrum.
	
\begin{figure}
		\centering
		\includegraphics[scale=0.21]{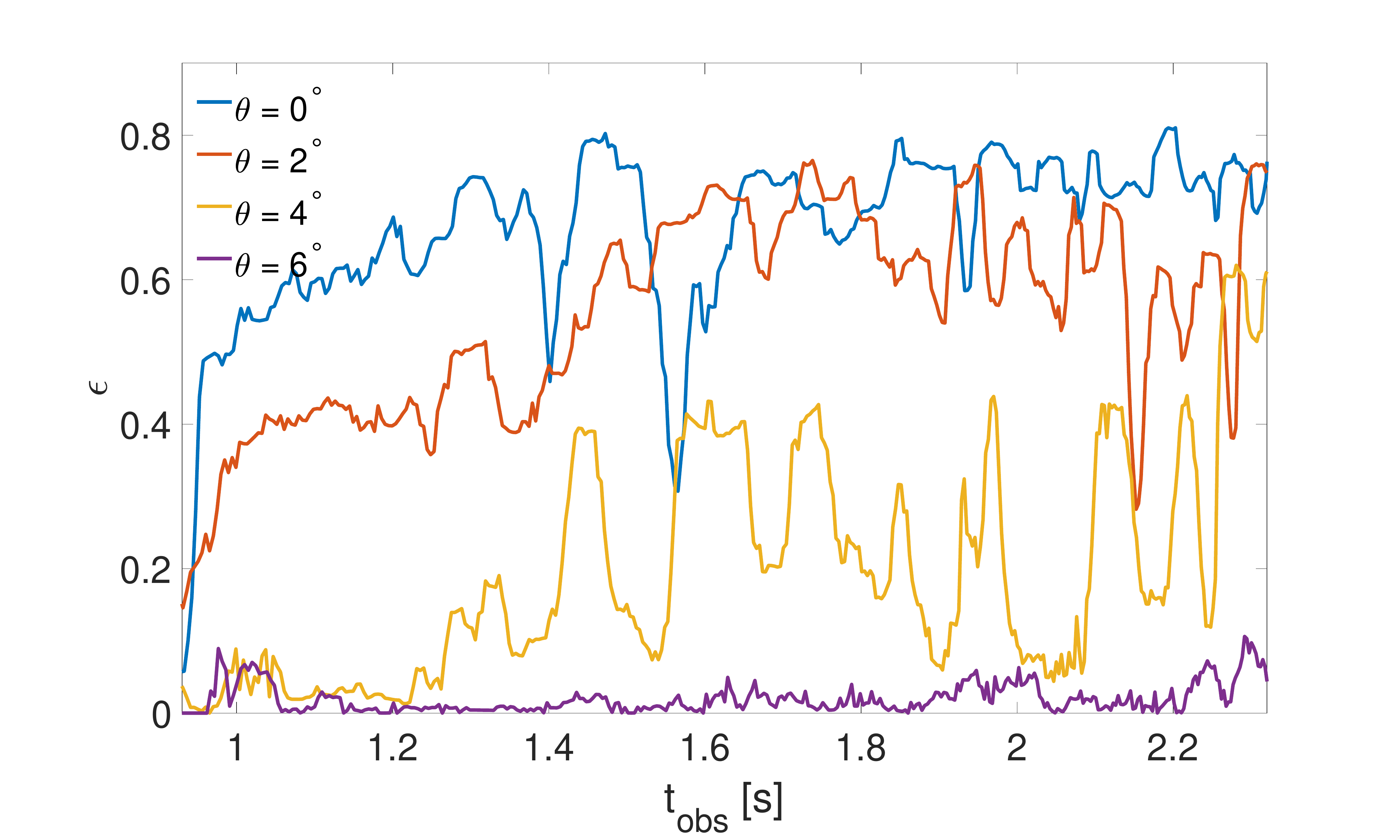}
		\caption[]{The radiative efficiency of the sGRB model $ \F $ at various angles. The re-opening angle is $ \theta_n \approx 3^\circ \approx 0.4\theta_0 $. The time $ \tobs = 0 $ corresponds to the time of the merger. The jet launch takes place at $ \tobs = 0.6\s $.
		}
		\label{fig:sgrb}
	\end{figure}

	Figure \ref{fig:sgrb} depicts the radiative efficiency for fluids at various angles in model $ \F $.
		While the on-axis efficiency is indeed higher, other characteristics are very similar to the lGRB models: The efficiency shows a clear evolution in time with large fluctuations over timescales of a few tens of ms.
		When considering the off-axis emission for our sGRB model, we find that the re-opening angle of the jet above the collimation shock is $ \theta_n \approx 0.4\theta_0 $. The angular dependency of the efficiency is also similar to lGRBs, with a roughly constant efficiency inside $ \theta_n $ followed by a sharp decline.
	
%%%%%%%%%%%%%

	\section{Comparison with previous works}
	\label{sec:previous}
	
	Jet simulations in 2D and 3D have been carried out previously by different
	authors. We turn now to highlight some differences between our results and other recent works.
	\citet{Lazzati2009} performed 2D axisymmetric simulation (also compared with a 3D setup) of jet propagation in a collapsar for a particular set of 
	parameters;  a jet with an opening angle  $ \theta_0 = 10^\circ $ and two-sided power $ L_j = 1.066 \times 10^{51}\erg~\s^{-1} $, injected into a $ 16\msun $ Wolf-Rayet progenitor star from a radius of $0.025 R_\star$
	\footnote{\citet{Lopez-Camara2013} repeated this calculation in 3D, but terminated the simulation early on, making a full comparison with their 2D run impossible.}.
	They found an efficient photospheric emission that they interpreted as caused by the presence of an oblique shock near the jet's head.  Their simulation  features a prominent sequence of collimation shocks that break out of the star following the jet,
	reaching large distances at the onset of the prompt emission phase, in marked 
	difference to our results.  We attribute this difference to the immense jet power adopted in their simulation and the dimensionality of the simulation.   
	Indeed, by repeating their experiment with the same jet power we find that the primary collimation shock breaks out after about 20 seconds, 
	both in 2D and 3D, in accord with their results.    As described in \S \ref{sec:simulations} above, at lower jet powers breakout of the shock is not expected 
	prior to emission (it takes at least 100 seconds for the shock to break out in models A,C-E) and the resulting structure above the star is different than that 
	reported in \cite{Lazzati2009}.   
	Note that the jet power adopted in \citet{Lazzati2009} implies an isotropic equivalent energy of $E_{iso}>10^{54}$ ergs, which is rare.
	Furthermore, while in 2D we also observed a sequence of relatively strong, repetitive collimation shocks, in 3D the 
	secondary shocks are nearly completely smeared out by mixing. 
	
	In general, axisymmetric 2D simulations have two fundamental differences from 3D. First, 2D jets tend to accumulate massive stellar material on top of their heads, unlike 3D jets.   This massive plug is an artifact of the imposed symmetry, that prevents the wiggling of the jet seen in 3D simulations, which 
	limits the amount of mass accumulated at the head.  Only jets having a sufficiently large opening angle (as in models C, D) exhibit a relatively massive plug in 3D.   Second, the mixing described in $ \S $ \ref{sec:mixing}, and seen in previous works (e.g. \citealt{Zhang2003,Rossi2008,Lopez-Camara2013}), is absent in axisymmetric 2D simulations. This mixing, which is triggered by instabilities developing along the jet-cocoon interface 
	in the collimation zone leads to additional, sporadic loading above the collimation shock, that renders the efficiency highly variable and angle dependent, as discussed in the preceding section.
	
	\citet{Ito2015} have presented 3D simulations of lGRB jets with parameters similar to our models, and found 
	that the primary collimation shock breaks out shortly after the jet's head.  
	However, in their simulations the jet is injected from a radius of about one quarter the stellar radius, which we suspected
	is the reason for the faster breakout of the jet and the shock, since this injection radius is larger than the radius of the primary collimation shock
	measured in our simulations for nearly the entire simulation time.
	In order to check this we repeated  the run of model A with the injection radius shifted from $10^{-2}R_\star$ to $0.25 R_\star$ 
	and found that the collimation shock indeed breaks out after 9 seconds. 
	To elucidate how the choice of injection radius affects the evolution of the system, we performed a convergence test in which we changed the injection
	radius keeping all other parameters the same.  We found convergence for injection radius smaller than about $0.03R_\star$.   
	We emphasize that, apart from its relevancy to the analysis outlined above, the resultant structure beneath the photosphere is of outmost
	importance for radiative transfer calculations that are based on snapshots of 3D RHD simulations \citep{Ito2015,parston2018,ito2018} (which are not included in our analysis).
	
	\section{Summary and Conclusions}
	\label{sec:summary}
	We performed a set of 3D numerical simulations of jet propagation for a range of GRB parameters that encompasses both long and short bursts, and used it to compute the minimum radiative efficiency of prompt photospheric emission along different sight lines. We find that the presence of a collimation shock strongly enhances the efficiency, despite substantial mixing of jet and cocoon material in the vicinity of the shock. The numerical results have been found to be in a good agreement with a simple analytic criterion derived in \S \ref{sec:analytic} for the critical load below which high efficiency is expected. A summary of our findings is as follows:
	
	\begin{itemize}

		\item  Mixing at the collimation throat leads to a substantial baryon loading downstream of the collimation shock, 
		and a consequent stratification of the flow.   The re-accelerating flow outside the star exhibits a quasi-uniform core
		of opening angle $\theta_n\approx 0.3\theta_0$, where $\theta_0$ is the jet opening angle at the injection point, and
		progressively heavier fluid on streamlines inclined at larger angles. 
		
		\item The radiative efficiency within the core is limited by baryon loading at the jet injection point, and is found to be very high, unless loading at the origin limits the asymptotic Lorentz factor to $\Gamma_\infty<100~ L_{iso,52}^{1/4}$.  Outside the core the efficiency declines with angles
		due to an additional loading by mixing. For example, in model $ \D $ a jet that is launched at the base with an opening angle of $\theta_0=14^\circ$, has a core of $\theta_n \approx 5^\circ$, where the average efficiency is 0.75 and along the sight line  $\theta=12^\circ$ it drops to about $10\%$.
				
		\item Even though the jet is injected continuously with a constant mass flux, mixing above the collimation shock  leads to a stratified outflow with highly variable baryon load. This in turn leads to 
		rapid spatial changes of the efficiency along the jet, that should give rise to a rapid variability of the observed prompt photospheric emission. We find variations over time scales of a few tens of milliseconds in the observer frame, however, it is unclear at present whether the fastest changes are actual  or limited by resolution.
		We also find an overall gradual increase of the efficiency over time, caused by diminishing mixing in our models.
		
		\item The variable baryon loading that the mixing generates near the collimation shock leads to large differences in the Lorentz factor of different fluid elements. This in turn leads to internal shocks between fast and slow elements, some of which take place below the photosphere.
		That can alter the observed spectrum significantly \citep{keren2014} and enhance the radiative efficiency. Our resolution prevents us from obtaining a reliable estimate of the energy that is dissipated in these shocks and we leave a more quantitative treatment of these shocks and their effect to a follow up paper.
		
		\item The observed color temperature (i.e., peak energy) of the photospheric emission is affected by photon generation above the collimation shock. 
		We find that, quite robustly, the temperature behind the collimation shock is regulated by vigorous pair creation, and is roughly 50 keV in the fluid
		rest frame.  Boosted to the observer frame, it is increased by a factor of $\sim \theta_0^{-1}$, consistent with observations.   This offers
		a natural explanation for the observed SED peaks.
		
	\end{itemize} 
	
	The main conclusion from our study is that for typical GRBs, a prominent photospheric component, likely dominating the prompt emission, cannot be avoided unless baryon loading is considerably higher than previously thought, or the jet remains magnetically dominated well above the photosphere. 
	
	Since the flow above the weak collimation zone is adiabatic, the photon number is conserved along streamlines. This implies that for emission above the coasting radius, where adiabatic cooling plays a role (expected at angles outside the jet's core), a linear relation between the inferred isotropic equivalent energy and the SED peak energy should hold, in difference from the Amati relation (but c.f., . \citealt{Lazzati2011,Lazzati2013}).  However, this argument ignores plausible effects caused by photon diffusion.  In particular, scattering of core photons sideways can conceivably dominate the off-axis emission.  
	A quantitative treatment of such effects requires full radiative transfer calculations.
	Such attempts have been reported recently \citep{ito2018,parston2018}, however, no convergence tests were performed and it is unclear how the numerical resolution of the RHD simulations affects the results.
	
	The temperature inferred from our analysis is in a good agreement with the observed peak energies of many bursts. Mild magnetization of the outflow is not needed to enhance photon generation, but may nonetheless have some effect \citep{Ludman2018}. In this paper we did not attempt
	to address the question whether the observed GRB spectra can form in hydrodynamic flows with a dominant photospheric component. 
	A naive expectation is that if the flow above the weak collimation shock indeed remains adiabatic all the way to the photosphere, the emitted spectrum should appear thermal (but not black body). However, we point out that the inhomogeneity of the flow (caused by the sporadic mixing mentioned above) can lead to a substantial broadening of the spectrum, owing to photon scattering off sheared matter \citep{ito2018,parston2018}. This process can modify some portion of the spectrum below the SED peak (though additional photon production may be needed) but may lack the energy needed to produce the high energy tail detected in many bursts.
	Those tails may be generated by sub-photospheric internal shock, if their efficiency is $\sim 10-20\%$, since such shocks tend to produce inherently broad spectra that mimic a Band spectrum \citep{ito2018b}. Whether the observed spectra can be explained by such processes remains 
	to be seen.
	
	\section*{Acknowledgements}
	
	We thank Davide Lazzati, Hirotaka Ito and Hamid Hamidani for useful comments.
	We also thank the anonymous referee for helpful comments that improved the presentation.
	This research was supported by the Israel Science Foundation (grant 1114/17). OG and EN were partially supported by an ERC grant (JetNS).	
	
	\bibliographystyle{mnras}
	\bibliography{Efficiency}
	
	\appendix
	\section{Semi-analytic profiles of an adiabatic fireball}
	\label{sec:numerical_calculation}
	
	In our post-process calculation, at each time we find the elements that are in the free acceleration phase. Then we infer their physical properties at future times by solving numerically their relativistic conservation equations.
	The conservation equations are given by \citep{Piran1993}:
	\begin{equation}\label{eq:conservation}
	\begin{cases}
	\frac{\partial}{\partial \tilde r} \big({\tilde r}^2{\tilde n}{\tilde \Gamma}\big) = 0 \\
	\frac{\partial}{\partial \tilde r} \big({\tilde r}^2{\tilde e}^{3/4}{\tilde \Gamma}\big) = 0 \\
	\frac{\partial}{\partial \tilde r} \Big[{\tilde r}^2{\tilde \Gamma}^2\big({\tilde n}+\frac{4}{3}{\tilde e}\big)\Big] = 0
	\end{cases}
	\end{equation}
	where $ {\tilde r} \equiv r/R_i $ is the location of the element, with $ R_i $ being the radius at which the element begins the conical expansion above the weak collimation zone, measured with respect to the shifted origin obtained as the focus of the radial trajectories (see Fig.\ref{fig:sketch}). The rest of the parameters are defined accordingly: $ {\tilde n} \equiv n/n_i $ is the number density normalized to $ n(R_i) $, $ {\tilde \Gamma} \equiv \Gamma/\Gamma_i $ is likewise the Lorentz factor normalized to $ \Gamma (R_i) $, and $ {\tilde e} \equiv e/e_i $ is the internal energy normalized to $ e(R_i) $, from which we find the pressure $ {\tilde p} = {\tilde e}/3 $.
	
\begin{figure}
		\centering
		\includegraphics[scale=0.21]{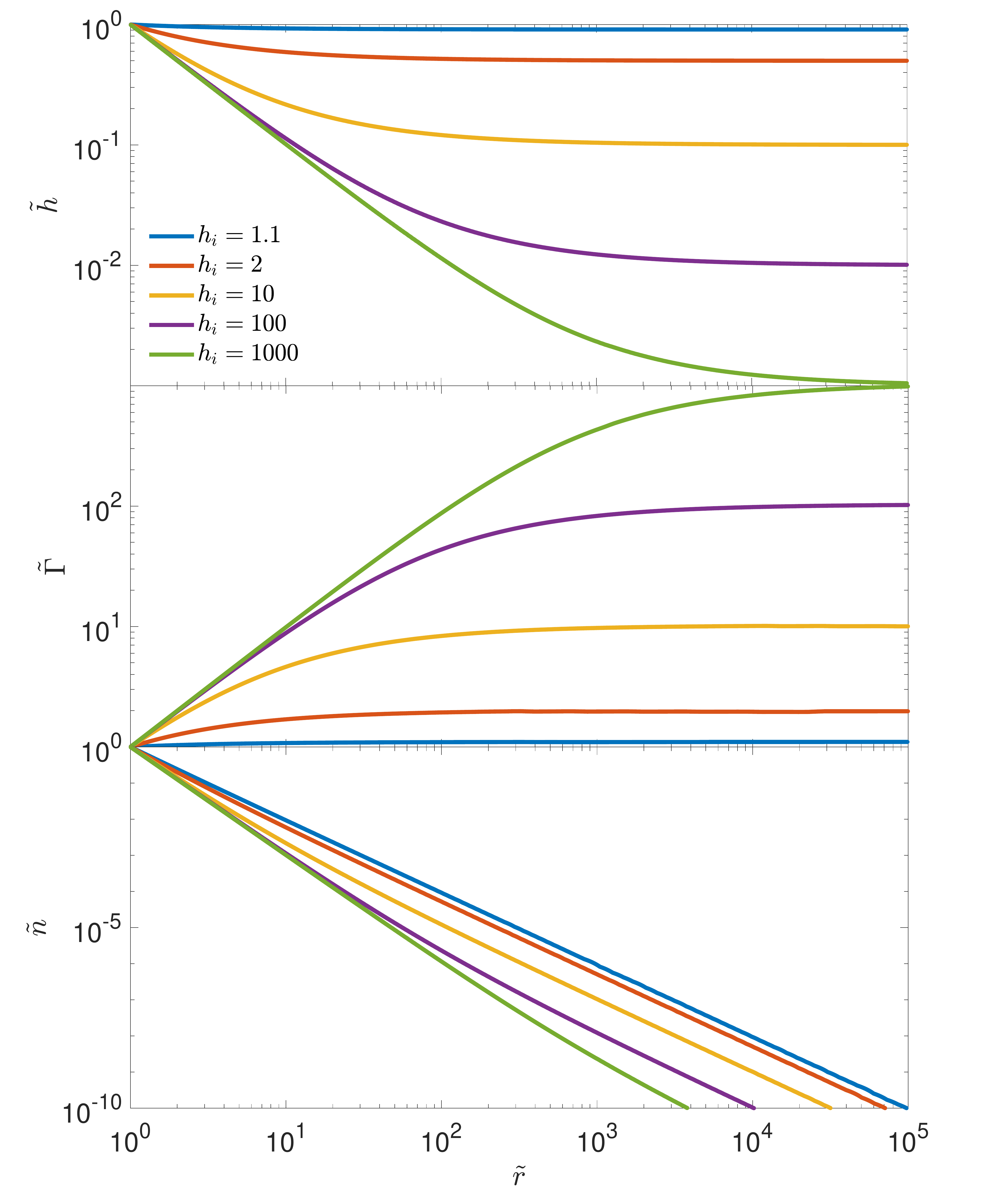}
		\caption[]{Universal profiles, obtained form the numerical integration of the
			adiabatic fireball equations \ref{eq:conservation}, for different values of the initial enthalpy per baryon $ h_i $.
		}
		\label{fig:ODE}
	\end{figure}

	Figure \ref{fig:ODE} displays universal profiles, obtained from the integration of Eqs. \ref{eq:conservation}, for different values of the initial specific enthalpy $ h_i \equiv 1+4{\tilde p}/{\tilde n}m_pc^2$. One can see the approximations in the free acceleration and coasting regimes. The former is prominent at early times for elements with large $ h_i $, where the relations $ {\tilde h} \equiv h/h_i \propto {\tilde r}^{-1} $, $ {\tilde \Gamma} \propto {\tilde r} $ and $ {\tilde n} \propto {\tilde r}^{-3} $ are satisfied. Elements with low $ h $ do not accelerate and are in the coasting regime from the very beginning. At late times, all elements are in the coasting regime, maintain: $ {\tilde h} \propto \rm const. $, $ {\tilde \Gamma} \propto \rm const. $ and $ {\tilde n} \propto {\tilde r}^{-2} $.
	
	\label{lastpage}
\end{document}